# Real-time Multi-instrument Autonomous Discovery of Novel Phase-change Memory Materials

Chih-Yu Lee [1], Haotong Liang [1], Ryan Kim[1], Austin McDannald [2], Carlos A Rios Ocampo[1,3], A. Gilad Kusne [1,2], Ichiro Takeuchi [1,4*]

[1] Department of Materials Science and Engineering, University of Maryland, College Park, MD, USA
[2] Materials Measurement Science, Division of the National Institute of Standards and Technology, Gaithersburg, MD, USA
[3] Institute of Research in Electronics and Applied Physics, University of Maryland, College Park, Maryland, USA
[4] Maryland Quantum Materials Center, University of Maryland, College Park, MD, USA
*Corresponding authors

## Abstract

Autonomous labs enable the integration of automated experiment execution, data analysis and decision making. The main challenge remains the integration of diverse data streams from multiple instruments, where the data is often heterogeneous and unsynchronized. The standard learning process of undetermined synthesis-process-structure-property relationships (SPSPR) usually relies on post-experiment analysis after data is fully collected, not during live experiments, and decision making is carried out independently across characterization equipment. Here, we demonstrate the Multi-instrument Autonomous Discovery (MAD) framework — combining structural property mapping and functional property optimization simultaneously in a closed-loop manner. As an example, we applied MAD to phase change memory (PCM) materials, and, in particular on the Mn-Sb-Te ternary, a previously unexplored materials system for PCM. A multi-output model is employed to merge data from x-ray diffraction (XRD) and electrical resistance measurements simultaneously through a co-regionalization kernel that models the relationship between them. The output probabilistic posterior and uncertainty quantification facilitate decision making with shared knowledge, while the goals are different across tasks. We aimed to maximize the knowledge of crystal structure distribution using non-negative matrix factorization (NMF), while in parallel, we find the composition with the maximum resistance value, an important figure of merit for PCM. Leveraging MAD, we found promising electrical PCMs and identified the SPSPR within 25 closed-loop iterations, corresponding to a seven-fold speed-up. The framework opens a new path of study in large-scale autonomous facilities, where future experiments can be run in parallel together, not independently, and the labs will not only make decisions for faster convergence, but also with better materials co-design criteria.

**Keywords:** *multi-instrument, machine-learning, active-learning, Bayesian optimization, autonomous experiment, thin-film, phase-change materials, materials optimization*

## 1. Introduction

Autonomous experimentation (AE) systems make informed decisions at each iteration based on data collected from automated instruments and learn from acquired data and explore previously hidden physical insights. This paradigm is revolutionizing materials science by fundamentally reshaping the pace and efficiency of optimization and discovery processes.[1–7] Exploiting high-throughput synthesis methods, such as thin-film combinatorial library, can open up vast design spaces across diverse compositions and processing parameters, generating large volumes of data for this purpose. [8,9] However, the bottleneck often lies in the considerable gap between synthesis and characterization, as complex physical property measurements can slow the translation of combinatorial data into physical models. To address this challenge, AE platforms are essential, enabling real-time decision-making and accelerating the path from data acquisition to insight.

Autonomous laboratories are typically implemented as single-instrument (and thus a single property characterization) closed-loop systems, in which an artificial intelligence (AI) model iteratively guides physical experiments through a single modality.[10] We refer to such an AI-instrument unit as an *agent*: an autonomous module that plans, executes, and interprets experiments within its own modality. Extending this paradigm to multiple heterogeneous instruments requires *orchestration* to coordinate interactions and enable information exchange among agents. As a result, experimentation shifts from a sequential paradigm to a parallelized, collaborative mode of operation. [11–14] Some open-source software frameworks, e.g., ChemOS[15], HELAO[16,17], and ESCALATE[18], featuring streamlined pipelines are designed to support orchestrated AEs. These systems typically require: (1) seamless instrument automation with robust handling of asynchronous experimental events; (2) integrated AI algorithms coupled with scalable computational infrastructure; and (3) reliable, bidirectional data communication between researchers and robotic agents. Notably, ChemOS 2.0 has demonstrated a complex materials discovery workflow by tightly integrating computational tools with experimental execution, enabling the identification of new materials for organic solid-state lasing devices.[19] As part of the Autonomous Formulation Lab framework, a "dueling robots" scheme has been enlisted to perform, small-angle x-ray scattering and small-angle neutron scattering, collaboratively and concurrently to study binary co-assembly of polymer and ceramic nanoparticles used in coatings.[20]

While *multi-agent* architectures will be a pivotal component of next-generation self-driving laboratories for addressing complex scientific challenges[14], the central objective of materials science remains the elucidation of synthesis–process–structure–property relationships (SPSPR). Learning SPSPR is fundamental to accelerating materials discovery and optimization, as it provides a blueprint for materials exploration, and particularly in condensed-matter systems, where enhanced properties often emerge (1) within narrow phase regions or (2) along phase boundaries. This objective can be systematically pursued using multi-task Gaussian process (GP) models[21], which learn shared covariance structures over input-dependent features to capture inter-property correlations. Such models enable principled information transfer across related tasks, leading to enhanced predictive accuracy, and have been successfully applied to alloy design involving multiple mechanical property objectives. [22,23] An alternative approach of SAGE[24] uses Bayesian coregionalization to represent the piecewise nature of phase-dependent material properties, merging knowledge across disparate data sources. While the co-orchestrated workflow of multimodal characterization tools, demonstrated using piezoresponse force microscopy and micro-Raman spectroscopy on a combinatorial Sm–$BiFeO_3$ library, successfully guided exploration of SPSPR in the work of Slautin *et al.*[25], the incorporation of learning directly within live AEs remains unrealized.

Combinatorial synthesis coupled with high-throughput characterization has emerged as an effective paradigm for accelerating materials discovery and design[9,26,27]. Deciphering the constituent phases, namely, constructing the crystal-structure phase diagram from diffraction patterns, is a critical step in formalizing SPSPR. However, this task is often costly and time consuming in practice. Rapidly charting the structural phase distribution across multi-component compositional phase spaces mapped on combinatorial libraries (i.e. composition spreads) is challenging, owing, for instance, to the prevalence of mixed-phase regions and the potential presence of non-equilibrium phases. A variety of algorithms have been developed to tackle the analysis of a large number of diffraction patterns to map structural phase diagrams.[28–32]; however, many of these approaches rely heavily on phase identification within supervised learning frameworks. Following the work of Long *et al.*[33], who demonstrated non-negative matrix factorization (NMF) as an unsupervised pattern de-mixing approach to addressing this problem, numerous extensions and variants have subsequently been developed.[32,34–36] They provide segmentation of compositional phase regions delineating phase boundaries within a given compositional space, enabling systematic study of SPSPR. For instance, the CAMEO[26] algorithm utilizes this knowledge to define phase boundaries in a piecewise function to fit and model functional property data. CAMEO uses a switching acquisition function strategy that first targets phase map knowledge and then switches to materials optimization. However, CAMEO's surrogate models for composition-phase and composition-property are independent GPs. The models for the functional property gain knowledge from the phase map, i.e., knowledge flows in one direction. Without a shared representation with knowledge flowing in both direction, the method does not fully capture or quantify inter-property correlations, limiting interpretability and generalization. SAGE improves on CAMEO facilitating simultaneous information sharing between the structure and property measurements, whereas its implementation within real-time AEs has not yet been demonstrated.

To the best of our knowledge, there is a lack of real-time data-fusion platforms that can jointly analyze correlations across multiple data streams and integrate this information into Bayesian optimization (BO)–guided orchestrated AEs are currently lacking. Developing such capabilities is critical for enabling adaptive decision-making and thus accelerating materials discovery, as it can bypass and speed up the traditional sequential measurement process, where analysis is deferred until datasets from disparate characterization results are fully consolidated. Herein, we report the successful demonstration of the Multi-instrument Autonomous Discovery (MAD), which enables parallel AE through BO built upon a shared and integrated knowledge foundation. In particular, we implemented the framework for an underexplored ternary thin-film system — Mn-Sb-Te (MST), within which there is a candidate magnetic topological insulator ($MnSb_2Te_4$) [37–40] and potential room-temperature magnetic PCMs. [40–43] The MAD enables joint prediction through the synergistic integration of (1) x-ray diffraction (XRD)-derived structural features and (2) electrical resistance measurements as a "*multi-task*" problem, employing distinct acquisition functions for each task within the active learning loop to realize a multi-objective optimization scheme. Continuous iterative data acquisition facilitates the construction of phase abundances and delineation of phase boundaries while electrical resistance contours are resolved across the ternary composition space. In addition, the MAD framework supports the reconstruction of XRD patterns and the quantitative elucidation of relationships between structural and functional properties. A single live run on MST simultaneously achieved phase mapping and materials optimization within 5 h, demonstrating superior efficiency compared to AE employing independent GPs and conventional time-consuming grid mapping, which usually takes several days. The proposed workflow is versatile, flexible, and modular, i.e., both hardware and software components are interchangeable, establishing a robust foundation for co-orchestrating multiple characterization instruments and extending AE into higher-dimensional, multi-instrumental environments.

## 2. Results

In thin-film libraries, multiple crystal phases often coexist, and these phases are closely linked to the material's functional properties. However, the presence of unknown and potentially metastable phases can complicate simulation and modeling workflows. On the experimental side, structural characterization via XRD becomes particularly challenging in such systems: overlapping peaks, compositional lattice shifts, and strong texture in thin films impede unambiguous phase identification and quantification. Moreover, XRD data acquisition remains time-consuming across entire libraries containing hundreds of different compositions, and thus it is pivotal to integrate structural analysis with functional property measurements to guide phase mapping. We perform experiments in parallel and with continuous live updating of property maps over the ternary diagram, as shown in **Figure 1-a**, which ultimately informs the selection of subsequent compositions to identify phase transition regions or phase boundaries. Prior to the live experiment, wavelength dispersive spectroscopy is first used to map the composition of the MST thin-film library, providing spatial coordinates for AEs. Two compositionally identical spreads are prepared by co-sputtering. A patterned physical mask was used to define 177 gridded square regions (each 4 mm x 4 mm), and so we obtain spatially addressable composition mapping on the library wafer. Subsequently, the two spreads are mounted on two remote instruments, namely, XRD and contact probe station (to optimize electrical resistance), which are connected to a central main agent via a server–client model over the computer networking, enabling command-driven data exchange and centralized computation. Contact measurement is intrinsically more time-efficient; the system manages this asymmetry by idling the faster process while awaiting completion of the slower one. Note that in this case, the speed is limited by diffractometer. However, the throughput bottleneck arising from instrument idling can be mitigated through a multi-agent implementation, as discussed later. In each iteration, the central controller, leveraging a centralized data repository and a multi-output GP model, performs joint inference across two property domains while identifying cross-property correlations.

To demonstrate the capability of this novel approach, we applied it to the AE–driven mapping of structural phase distribution and the resistance of MST thin-film libraries, spanning a wide compositional range with constituent elements varying from 7 % to 78 %. **Figure 1-b** is a photo of an experimental setup, where two compositionally identical spreads are placed on separate instruments: one is annealed to its crystalline state for XRD, and one is as-deposited as amorphous state for electrical measurements. In this work, we aim to optimize materials with the highest electrical resistance in amorphous states ($R_{\mathrm{amo}}$), although the electrical resistance contrast between amorphous and crystalline states ($\Delta R = R_{\mathrm{amo}} - R_{\mathrm{cry}}$) is more commonly used to evaluate PCM performance. Given that the amorphous structure acts as a precursor, influencing the structure adopted upon crystallization[45], $R_{amo}$ provides valuable information of materials in amorphous state, which will later be merged with structural information measured in crystalline state. For this reason, we take $R_{\mathrm{amo}}$ as the figure of merit, that is, maximal electrical resistance on the amorphous composition spread library. The diffractometer automatically performs three-frame scans in 2θ ranges of 11°-28°, 27°-42°, and 42° -57° , and data integration is carried out using custom Python-based software, some of which are provided on **GitHub**. The probe station conducts three independent measurements to provide statistically meaningful two-terminal resistance measurement results at each separated composition region. The recording results of a live experiment is included in **S1**. Hardware control, data communication and computation codes are available on **GitHub**. We note that the active learning campaign was initialized with 5 randomly selected points on the library wafer. During the live AE, data is independently collected from the two tools and transmitted to the central computer, where joint prediction is first conducted. The subsequent BO decision-making step is performed independently for each instrument.

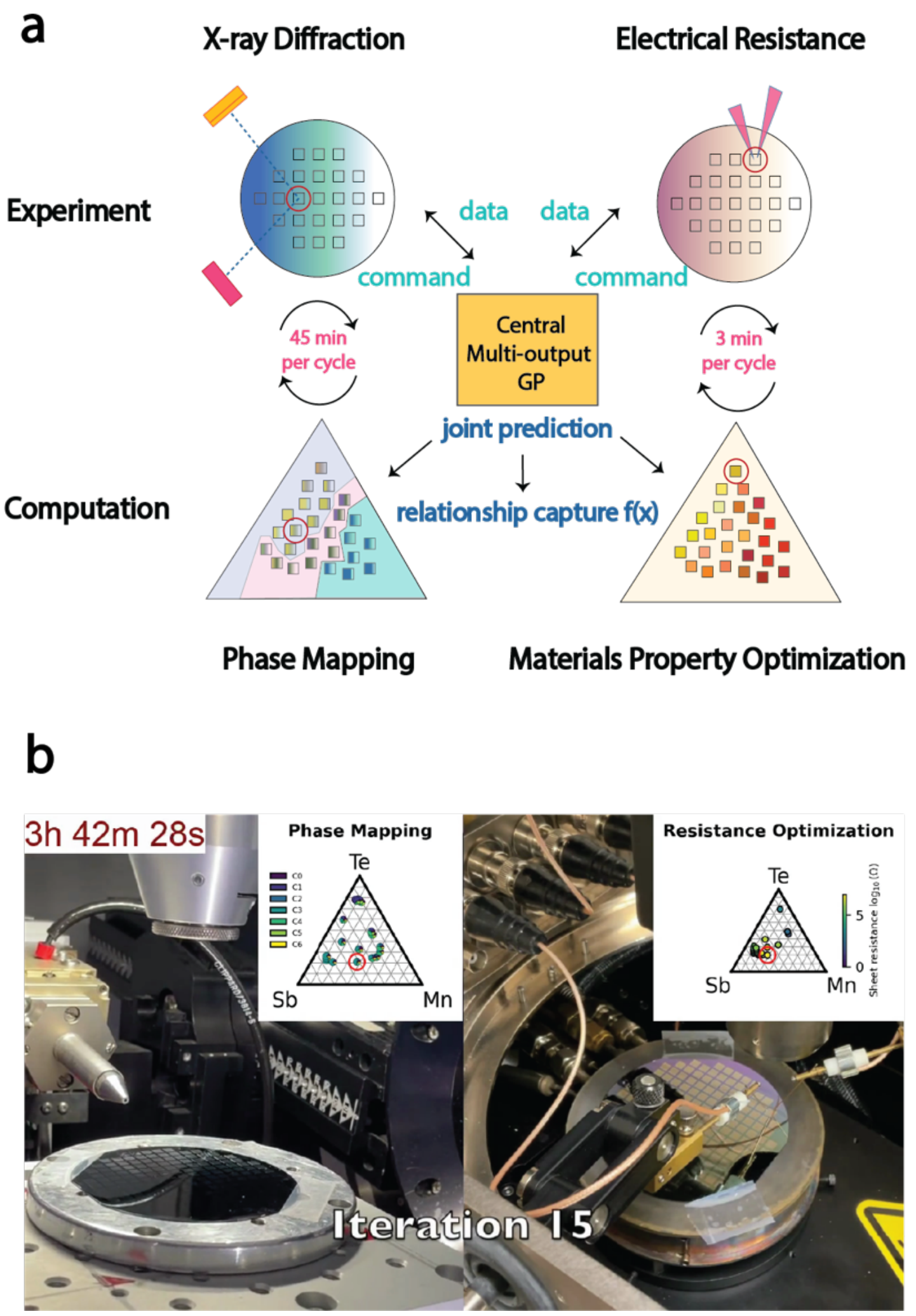

***Figure 1 Overview of the MAD workflow.*** (a) Synchronized operation of two spatially separated instruments, communicating via the campus internal network. A central computer acts as the computational hub, receiving data and sending control commands to the local instruments. Most importantly, the central computer captures the real-time relationship $f(x)$ between structural and functional properties. (b)Snapshot of the live experiment. Insets show real-time data acquisition and decision-making represented on a ternary diagram. The video recording is provided in S1.

Given that multiphase regions naturally exist in compositional phase maps, nonnegative matrix factorization (NMF) provides a promising unsupervised learning approach that decomposes XRD pattern data into a reduced set of nonnegative components and corresponding weights, with the latter representing the abundance of each structural phase (end member) at a given composition.[32,36,46] Most importantly, NMF converts the clustering problem into a continuous regression form that can be directly incorporated into a predictive model. In MAD, diffraction patterns (the original experimental diffraction is represented as a matrix, where each row is a diffraction pattern, and each column corresponds to intensities at a particular 2θ value) were first pre-processed by background subtraction and subsequently factorized into seven end members (basis pattern matrix) and their corresponding abundances (coefficient matrix denoted as C0-C6), as shown in an example in **Figure 2-a**. The number of end-member components, *N*=7, was selected based on the number of crystalline phases in a ternary system based on a reasonable and simple assumption/guess — that there might be three elemental phases, three binary compounds, and one ternary compound. As demonstrated in **S2**, this choice of *N*=7 is sufficient to minimize the reconstruction error in NMF, as verified by the explained variance elbow test. In **Figure 2-b**, each individual pie chart represents the phase makeup for a specific composition spot. The phase abundances were concatenated with the resistance measurements to form an eight-dimensional dataset with each measured composition entry including 7 phase abundance coefficients (C0-C6) and resistance which was subsequently used as input for the coregionalized GP for joint prediction, providing both the predictive mean and associated uncertainty. It is noteworthy that in NMF, the abundance representation intrinsically captures the mixed-phase character of the system. This contrasts with label-based approaches[47], which treat single-phase and mixed-phase regions equivalently and independently, thereby overlooking the compositional continuity inherent to phase coexistence. Using the predicted abundances, the corresponding diffraction patterns can be reconstructed through the matrix product with the NMF-derived basis pattern matrix. Note that NMF is performed on the available data at each iteration; therefore, the basis pattern matrix evolves as new data points are added. The red pattern in **Figure 2-a** shows the reconstructed XRD while the black one is the experimentally measured pattern. The algorithmic workflow of MAD, illustrated in **Figure 2-c**, indicates that XRD reconstruction, as a byproduct, provides an indirect means to assess the performance of the coregionalized GP. In addition, the XRD latent representation, regarded as another byproduct of the model, encodes the underlying structure of the input space.

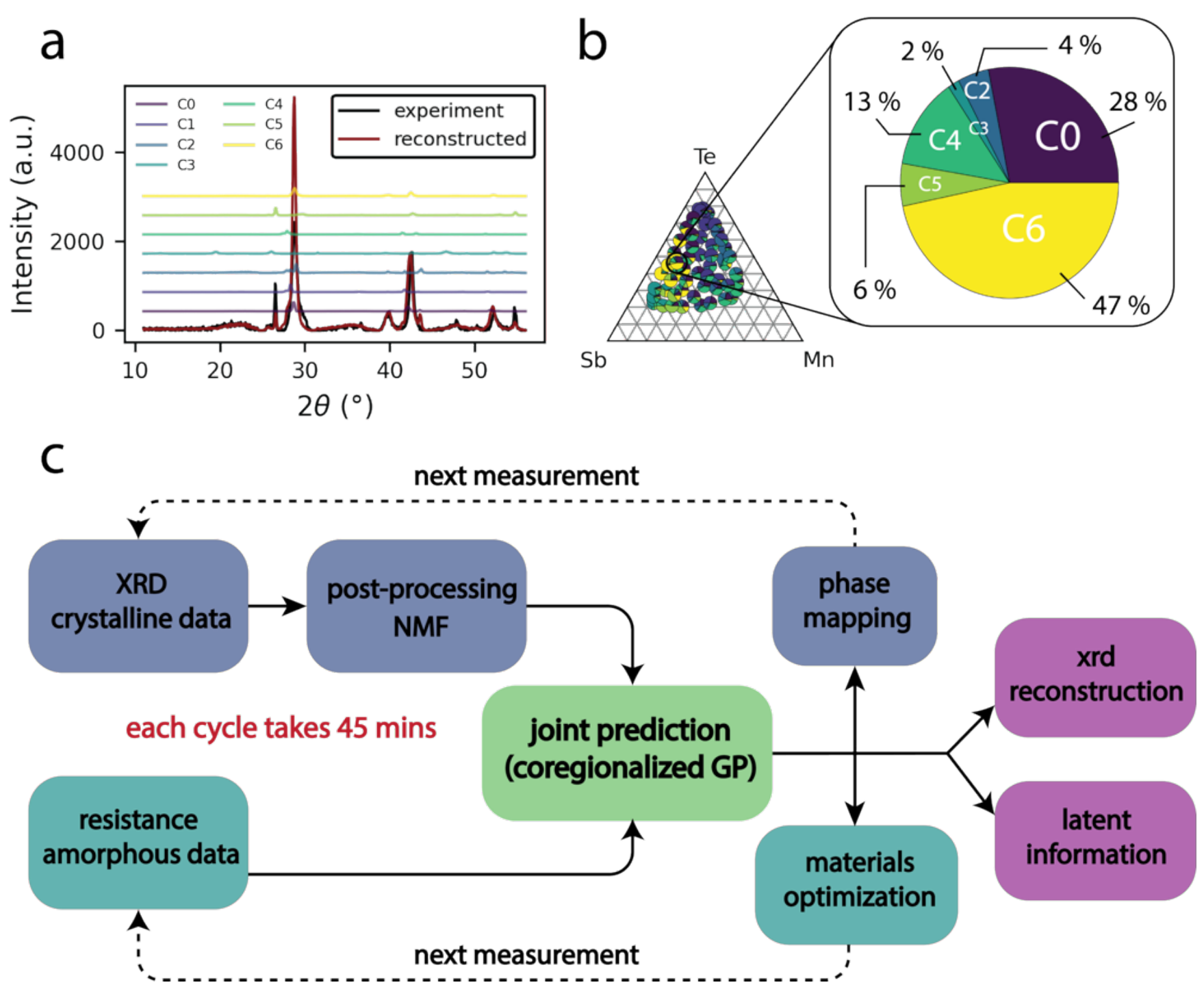

***Figure 2 NMF decomposition and algorithms of the MAD.*** (a) Experimental XRD pattern and its inverse reconstruction using NMF, based on basis components C0–C1 and abundances predicted by GP. Noted that the basis component are displayed as a cascaded waterfall plot, so the intensity axis is arbitrary. (b) Phase configurations represented as pie charts, where each chart illustrates the proportion of individual NMF components within a sample.

(c) Algorithmic workflow of MAD. Two tasks perform data collection independently, followed by joint prediction and separate decision-making in each iteration.

Following joint prediction, the interrelationships among the data streams can be extracted from the GP Intrinsic Coregionalization Model (ICM) coregionalized kernel matrix, which is defined as:

$$k_{ICM}\big((x,d),(x',d')\big) = B_{dd'} \cdot k(x,x')$$

where

$$B = WW^T + diag(\kappa), W \epsilon\ \mathbb{R}^{D\times R}, \kappa\ \in \mathbb{R}^{D}$$

Here, $x$ denotes the input vector in composition space, and $x'$ represents another input location. The function $k(x,x')$ is the spatial kernel that quantifies similarity between two input points in the composition space. The index $d \in \{1,\ldots,D\}$ denotes the output channel (task), corresponding to either one of the seven phase abundances or the resistance measurement, and $d'$ denotes a second output channel. Accordingly, $B_{dd'}$ represents the $(d,d')$-th element of the coregionalization matrix $B$, which encodes correlations between output channels $d$ and $d'$. Here, $D$ denotes the total number of outputs (in this case, 8: seven phase abundances and one resistance), and $R$ represents the latent rank of the coregionalization structure. In this study, we set $R = 1$, corresponding to a rank-one intrinsic coregionalization model, for parsimony and identifiability in the low-data regime, reflecting the shared dependence of all outputs on composition; $diag(\kappa)$ preserves per-task residual variance and prevents over-coupling. Note that dimensional consistency across heterogeneous input variables is ensured through normalization, as detailed in the Methods section. $W$ is the latent mixing matrix, and $\kappa$ represents output-specific variances. Both parameters were optimized when training the GP model. The matrix $B$ quantifies the covariance between different outputs, i.e., structural properties and the functional property, while the correlation matrix between outputs can be obtained by normalizing $B$. The latent matrix $W$ can also be retrieved to examine the low-rank structure underpinning multi-output correlations. By constraining $B$ to a diagonal form—effectively suppressing cross-output covariances—the model reduces to a set of independent GPs, in contrast to the coregionalized formulation that explicitly captures shared structure across outputs. We used it to benchmark and assess the benefits of the coregionalized formulation that explicitly captures shared structure across outputs. Tracking these correlations throughout the closed-loop process is inherently challenging, as the basis diffraction pattern matrix evolves with each iteration. Nevertheless, the evolving correlation structure reflects the relationships between the resistance mapping and each NMF-derived component, which may not directly correspond to distinct crystalline phases. In the animation **S3**, we see the phase-resistance correlation dynamically evolves through the iterative experiments, which provides insights of SPSPR in real time.

Following joint prediction from the coregionalized GP, acquisition functions were employed to balance exploration and exploitation, thereby guiding the selection of subsequent experimental conditions. The decision-making processes for the two tasks were independent, as each pursued a distinct objective: (1) maximizing phase knowledge through exploratory sampling for XRD, and (2) identifying the material composition with highest electrical resistance through exploitative optimization with probe station. Accordingly, maximum uncertainty (MU) across all predicted phase abundances was used as the acquisition function for exploratory sampling, while expected improvement (EI) was applied to iteratively balance exploration and exploitation during optimization. Previously measured compositions were excluded from consideration, and the compositions with the highest MU or EI were selected as the next experimental targets (see **S4** for detailed equations and decision pathways).

After completing the full AE campaign (i.e., collecting XRD and $R_{amo}$ for all compositions), we treated analysis of the resulting dataset as the ground truth. Using this reference dataset, we subsequently evaluated the performance of the models in a post-experimental analysis. The accuracy of phase mapping was evaluated by comparing reconstructed XRD patterns with experimental measurements using the cosine similarity score (CSS) [48] and dynamic time warping (DTW) distance[49]. CSS quantifies overall pattern alignment, while DTW accounts for peak shifts, providing a robust assessment of reconstruction quality. Note that the reconstruction error represents a convolution of the NMF and multi-output GP components (i.e., reconstruction uses the multi-task GP model's computed XRD abundances); for simplicity, we assume the NMF to be ideal, as discussed further in the *Discussion* section. Functional property optimization performance was evaluated using the percent error and minimum regret distance (MR). Percent error quantified the overall prediction accuracy, while MR measured the deviation between the best measured composition and true global maxima, reflecting the likelihood of selecting suboptimal compositions (see **S5** for associated equations). **Figure 3** illustrates the evolution of the optimal-point search process from 0th to 177$^{th}$ iterations, i.e., full library mapping for this composition spread containing 177 spots. On one hand, with approximately 42 measurements, MAD succeeded in reconstructing XRD patterns with 95 % of similarity, implying that the algorithm gained sufficient phase knowledge. On the other hand, it took MAD fewer than 25 data points to find the optimal material (highest $R_{amo}$), with the percent error decreasing steadily and MR eventually converging to 0 and remaining stably there.

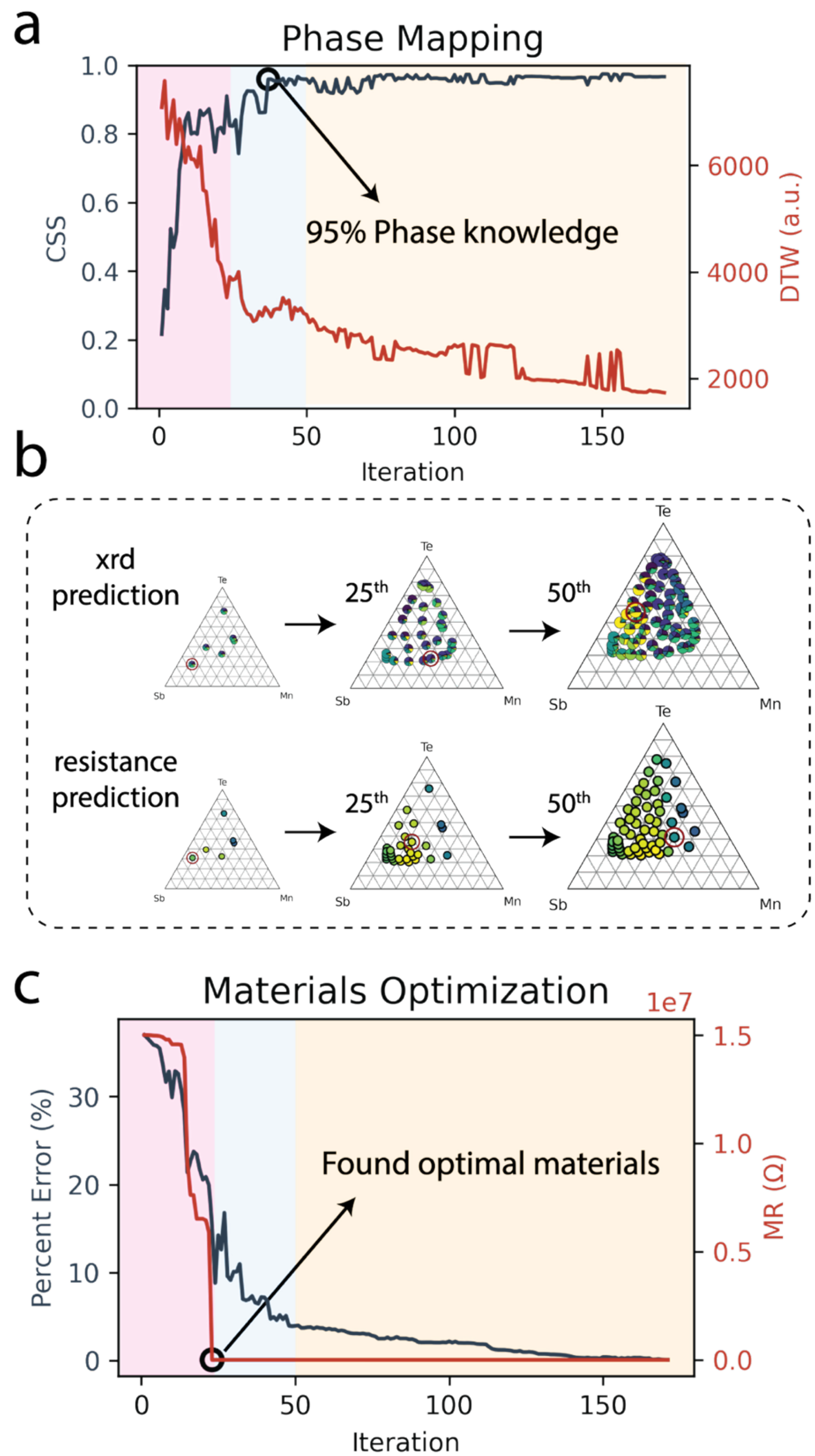

***Figure 3 Evolution and performance of iterative materials discovery.*** (a) Structural reconstruction accuracy during the discovery process, evaluated using cosine similarity score (CSS) and dynamic time warping (DTW) between reconstructed and measured XRD patterns. (c) Optimization performance as a function of iteration number. Left axis: percent error in the predicted optimal resistance relative to the ground truth. Right axis: minimum regret distance (MR), defined as the absolute difference between the true global maximum resistance and the best measured resistance among sampled compositions (units: Ω). (b) Evolution of the predicted property landscape at iterations #1, #25, and #50. Red circles denote experimentally measured compositions at each stage, illustrating the adaptive sampling trajectory guided by the model.

To benchmark the performance of MAD, we conducted *in silico* experiments using 10 sets of randomly initialized seeds for the starting points. We systematically varied (i) the GP kernel used for joint prediction, i.e., comparing the previously introduced coregionalized kernel with two independent kernels, and (ii) the decision-making strategy, i.e., contrasting BO based on our acquisition functions with random sampling. As shown in **Figure 4-a**, the BO-guided coregionalized kernel outperforms all other configurations, achieving the highest reconstruction accuracy with the fewest sampled data points. In contrast, conventional random sampling fails to converge even after 100 measurements. Moreover, the BO-guided approach exhibits improved stability, evidenced by its narrower performance distribution, indicating greater consistency in decision-making across different initializations. The introduction of a shared latent space enhances phase mapping. As illustrated in **Figure 4-b**, models employing the coregionalized kernel achieve consistently lower DTW scores, indicating better matching, compared to those using independent kernels, irrespective of the sampling strategy. With respect to materials optimization, the coregionalized kernel again demonstrates faster convergence compared to the no–shared-knowledge (independent kernel) scenarios, as shown in Figures **4-c**. Importantly, the acquisition function employed here explicitly balances exploitation and exploration, while also accounting for the uncertainty propagated from the concurrently learned phase information. Under this framework, the BO-guided coregionalized kernel exhibits a markedly slower reduction in predictive accuracy at the earlier stage, relative to its random sampling counterparts. In **Figure 4-d**, we further observe the robustness of the BO-guided coregionalized GP: once the MR distance reaches zero, it remains stable and does not deviate in subsequent iterations. Ultimately, it identifies the optimal material compositions using the minimal number of measurements, highlighting the efficiency gained from shared latent representations and uncertainty aware decision-making.

By employing a coregionalized GP, the closed-loop AE workflow is substantially accelerated relative to the independent GP framework. Moreover, MAD achieves at least a fivefold speed-up compared with conventional random sampling, effectively reducing approximately three full days of experimental time and associated resource expenditure. Because MAD leverages shared information across tasks, it helps bridge the discrepancies arising from measurements performed on different instruments and from problems of varying complexity. In such multi-task settings, tasks associated with simpler objectives tend to converge more quickly, while others may struggle to achieve comparable performance. By incorporating a shared latent structure, MAD effectively balances the two tasks, with both converging using approximately 40 data points. This coordinated learning leads to more uniform convergence behavior and enhances overall optimization efficiency. **Figure 4-e** schematically illustrates the different knowledge-sharing paradigms and highlights the importance of cross-task information exchange. In the conventional setting, no latent information is shared between structure and function; consequently, the intrinsic relationship between the phase diagram and the functional property landscape is neglected. Structural evolution and functional responses are modeled independently, preventing coordinated learning. In the CAMEO framework previously developed[26], structural mapping is performed first, and the kernel can be switched once sufficient phase knowledge has been acquired. However, this approach implicitly assumes that each phase region exhibits distinct and uniform functional behavior. Structure is effectively treated as a discrete label, which limits the ability to capture mixed-phase regions and gradual transitions across compositional boundaries. In contrast, MAD enables full knowledge exchange; thereby structural and functional measurements are jointly learned within a coregionalized GP. Moreover, by representing crystal structures as continuous multiphase abundances rather than discrete labels, MAD is able to recognizes that functional change boundaries may be tied to the presence of certain phases. This results in smoother, more physically consistent predictions of functional properties across composition space and improves both convergence and robustness in closed-loop optimization.

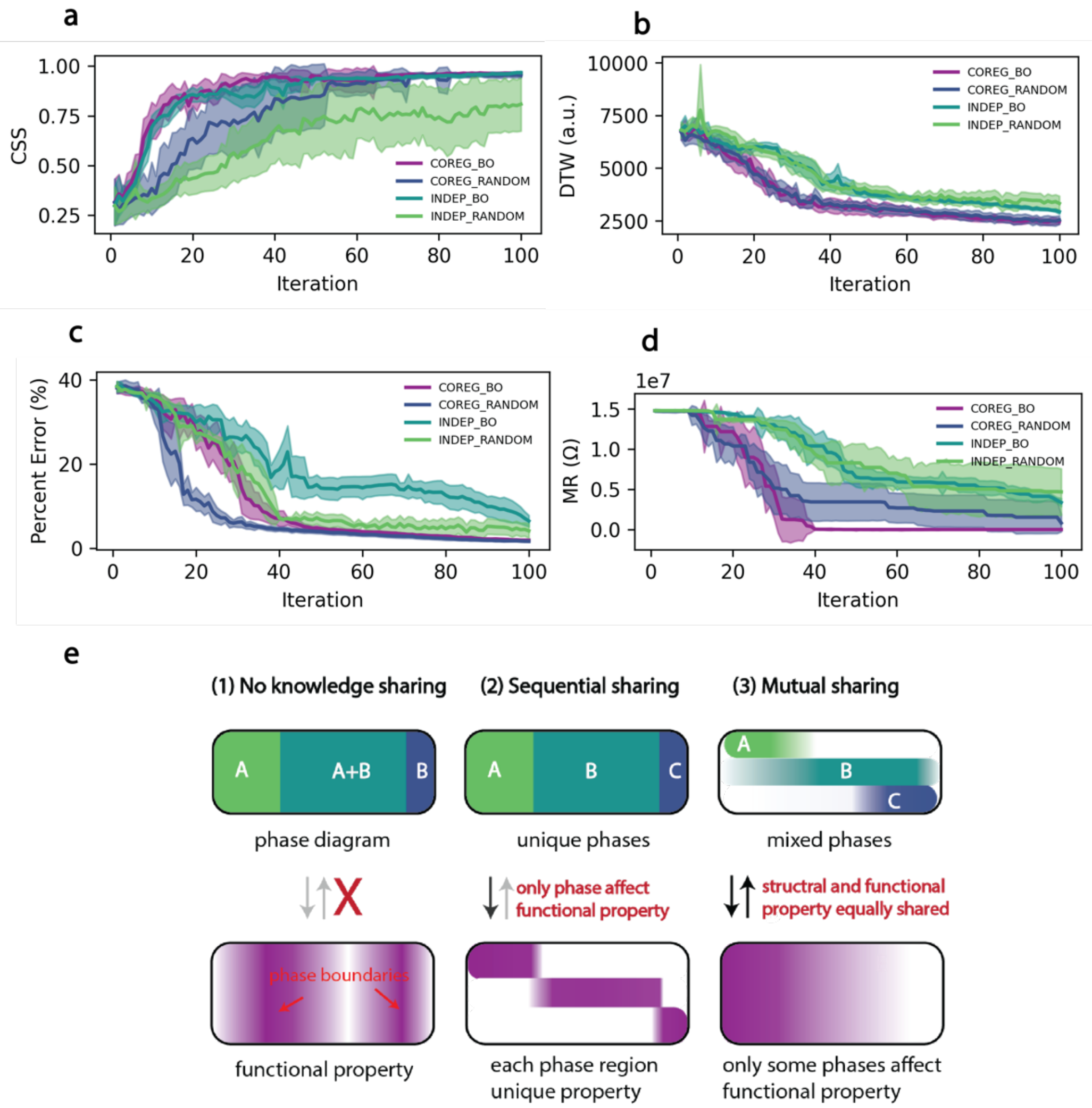

***Figure 4 Performance of different GP kernels, decision-making, and knowledge sharing strategies.*** Benchmarking the convergence speed and model stability as *in silico* AE proceeds. COREG_BO: coregionalized GP+ Bayesian optimization. COREG_RANDOM: coregionalized GP+ random sampling. INDP_BO: independent GPs+ Bayesian optimization. INDEP_RANDOM: independent GPs+ random sampling. Phase mapping performance is quantified through CSS (a) and DTW (b), whereas materials optimization performance is evaluated using percent error (c) and MR (d). (e) Illustration of different knowledge-sharing schemes across structural and functional properties.

To map the entire compositional library and fully demonstrate the acceleration enabled by the implementation of MAD in a live setting, AE was executed until all data points were sampled. The resulting phase configurations are shown in **Figure 5-a**. A transitional region characterized by a high fraction of the C4 phase emerges near the center of the ternary space. From the resistance mapping in **Figure 5-b**, we observe that Mn concentration is positively correlated with conductance, while the Sb-rich corner also displays elevated conductance despite its relatively low Mn content. A pronounced reduction in resistance $R_{amo}$ (four orders of magnitude) is observed with varying Sb–Mn concentration, largely independent of Te content. In contrast, the $R_{cry}$ remains uniform and predominantly conductive across the compositional space (see Supplementary Section **S6**), with measurements acquired post-AE on the crystalline spread. Our analysis identifies $Mn_{28}Sb_{52}Te_{20}$ as the optimal composition based on maximum $R_{amo}$ , while $Mn_{14}Sb_{43}Te_{43}$ yields the largest resistance contrast $\Delta R$ (approximately 5 orders of magnitude). We note that the discrepancy between the two resulting $\Delta R$ is minimal, and so the two compositions are considered potential optimal PCMs. The primary objective is to uncover the underlying correlations between structural features and functional properties, and the specific optimization target can be readily substituted within the same MAD framework.

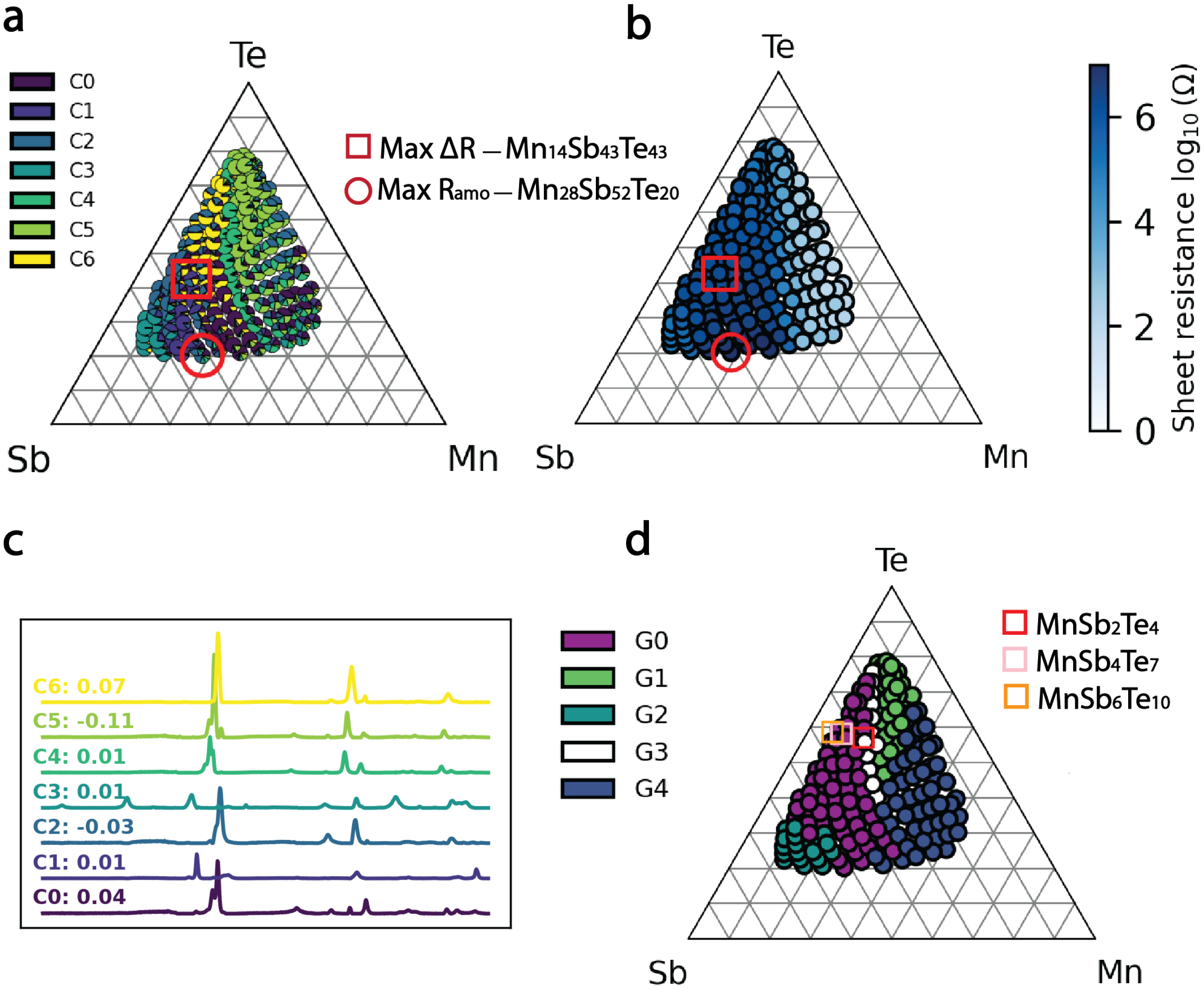

***Figure 5 Structural and functional property mapping of the entire thin film library.*** (a) Phase configurations represented in pie charts for each composition. (b) Sheet resistance mapped across the amorphous thin-film library $R_{amo}$. (c) Basis components derived from NMF at the final iteration of AE, i.e., GP trained with all data points; the numbers indicate the correlation between each component and resistance $R_{amo}$. Each component is represented as a profile with 2θ as the x-axis and intensity as the y-axis; however, the spectra are vertically offset for visualization, and absolute intensity values are not directly comparable. (d) Piecewise clustering of all collected diffraction patterns, with clusters indicated by color. Some well-known compositions are indicated with squares on the ternary diagram, while the remaining points correspond to compositions not yet explored.

Leveraging the coregionalized matrix, we extracted the correlation between each phase component and the measured resistance. The correlation values were computed using the equations provided in **S7**. As shown in **Figure 5-c**, the C5 phase (green) exhibits a relatively stronger association with conductive behavior, as indicated by their inverse correlation, whereas the C6 phase (yellow) is positively correlated with $R_{amo}$ and predominantly present in regions that remain relatively insulating. Despite their weak correlation with resistance, C3 and C4 are primarily distributed in the Sb-rich and transitional phase regions, respectively. It is important to note that the correlations are not directly comparable across iterations, as the NMF decomposition is performed independently at each step. Once the model stabilizes, these correlations can be analyzed systematically, enabling a more meaningful interpretation of the underlying trends. We also clustered the diffraction patterns using unsupervised learning with $M = 5$ clusters and found that the resulting phase boundaries align well with the NMF-based phase mapping. Several studied compositions[37,39,50] are concentrated in the G0 (purple) cluster, which exhibits a strong correlation with the C6 phase (**Figure 5-d**). We note that MAD is particularly advantageous for problems with smoothly varying functional properties that are not easily distinguished by structural features, as it allows phase mapping to be treated in a regression framework. In contrast, piecewise clustering as in SAGE[47] imposes a strong assumption that each cluster exhibits distinct and independent behavior, implicitly treating each phase region as equally contributory to the functional property. The assumption may not hold in practice, e.g., in the case of a continuous compositional spread. The fundamental difference between two approaches is discussed later.

According to Liu *et al.*, $MnSb_2Te_4$ crystallizes with a trigonal structure ( rhombohedral $R\bar{3}m$ space group), isostructural with the well-known $MnBi_2Te_4$.[51] It forms a natural homologous series described by $(\mathrm{MnSb_2Te_4})(\mathrm{Sb_2Te_3})_n$ with $n = 0,1,2, \ldots$, arising from different stacking sequences of septuple layers and quintuple layers along the c-axis. [52] Despite the mixed-phase character of the system, we performed further phase matching across the vast compositional space, 95 % of which has not been previously explored. We find that C0 and C2 can be indexed to the trigonal $MnSb_2Te_4$ structure, with systematic peak shifts. C4, C5, and C6 are associated with the trigonal Te phase, whereas C1 and C3 do not match any known reference structure, see **S7.** These results suggest that the trigonal $MnSb_2Te_4$ structure is central to the observed phase-change behavior, although a detailed assessment of competing effects from impurity phases remains necessary. This material system exhibits complex magnetic behavior, as Mn–Sb site mixing can modulate the stacking stability of layered structures and consequently influence the balance between ferromagnetic and antiferromagnetic ordering.[40,51] Nevertheless, the present ML-assisted study gleans the trend and narrows the candidate pool for potential PCM compositions in MST system, thereby providing a focused platform for future investigations into magnetic properties and the possibility of reversible phase-change switching.

## 3. Discussion and Future Work

NMF compresses high-dimensional XRD patterns into a small number of interpretable features, reducing GP input complexity and avoiding overfitting. Additionally, NMF acts as denoiser, which improves GP prediction stability and accuracy. Most significantly, it has the capability to capture mixed phase systems, enabling phase diagram reconstruction and smooth interpolation across compositions, which additionally manages to resolve overlapping diffraction peaks through the disentanglement of overlapping contributions. It is worth noting that advanced NMF techniques, such as Shift NMF[53], Sparse NMF[54] and Graph-regularized NMF[55], provide platforms to resolve peak shifting due to composition, strain, or solid solution effects, and improve robustness. In addition, future work can incorporate an NMF framework that automatically determines the optimal number of components, allowing model complexity to adapt dynamically to the evolving dataset. As MAD is dedicated to demonstrating the entire joint prediction pipeline, a standard NMF was exploited here.

A full Bayesian model can be achieved by a combination of Bayesian NMF[56,57] and heteroscedastic GP[58,59], both of which quantify uncertainties, enabling it to propagate through joint prediction and hence probabilistic phase diagram reconstruction with uncertainty quantification. However, this requires redefinition of coregionalized kernel matrix and adds complexity to the model, and thus we leave such implementations for future studies. We note that MAD assumes that the NMF decomposition is ideal, with uncertainty arising solely from the coregionalized GP. While GP uncertainty can indirectly influence XRD inverse reconstruction, MAD successfully achieves reconstruction errors below 10 % with just 30 measured data points. The current approach evaluates the performance of phase mapping indirectly, by comparing the similarity between measured diffraction patterns and predicted reconstructed patterns. However, a more direct and quantitative metric is needed.

Conventionally, phase mapping is typically formulated as predicting discrete labels. However, when categorical outputs (which require a non-Gaussian likelihood) are combined with continuous regression inputs, the resulting joint posterior becomes analytically difficult. Although Kusne *et al.* demonstrated a shared-likelihood framework in SAGE[47], the unified treatment of mixed likelihoods remains non-trivial. A model with shared latent GP but task-specified heads (heterogeneous likelihoods)[60,61] or multi-task variational GP with mixed likelihoods[62] can potentially address the problem despite convergence instability and complex hyperparameter tuning. MAD is advantageous, particularly when (1) a system has functional properties that exhibit smooth surface and thus lacks abrupt behavior change at the phase boundaries or (2) multiple coexisting phases, as we see in **Figure 4-e**. MAD with a coregionalized kernel outperforms multiple independent GPs in time and accuracy by leveraging shared information. However, this benefit is not universal, as correlations among tasks may be weak.

In fact, MAD builds upon SAGE[47] by extending the co-regionalization framework from discrete phase labels to continuous phase abundances, enabling joint modeling of structural and functional information. We posit that phase abundances provide a more expressive and physically meaningful representation than discrete labels, particularly for electrical and thermal conductivity. However, the advantage may be less clear for properties with strong phase-specific or symmetry-dependent responses, such as piezoelectricity or magnetoelectricity. The choice between these approaches largely depends on domain knowledge of the SPSPR. Notably, both frameworks are flexible and can accommodate either discontinuities or smooth variations; however, their effectiveness ultimately may hinge on sufficient sampling, particularly near phase boundaries, to accurately capture these transitions.

While MAD distributes time uniformly across tasks, simpler problems may still be forced to idle while more complex ones converge. This bottleneck can be mitigated by leveraging synchrotron beamline XRD, whose significantly higher photon flux enables substantially faster data acquisition compared to conventional laboratory XRD. In addition, the current architecture relies on a single central kernel, i.e., a single agent coordinating multiple instruments, which enforces an equal number of data points across tasks. The resulting throughput bottleneck due to instrument idling can be alleviated by distributing control across independent instrument servers coupled to a central AI server. In this multi-agent framework, each instrument operates asynchronously while still enabling shared learning across tasks.[13,63,64]

## 4. Conclusion

In summary, MAD establishes a parallel AE platform that departs from conventional serial phase mapping by explicitly addressing mixed-phase behavior and synchronously enabling phase identification and materials optimization. During live experimentation, MAD achieves stable, high-accuracy predictions of XRD patterns and functional properties, leading to an approximately five-fold acceleration in throughput. By leveraging a coregionalized Gaussian process model to dynamically interpret the SPSPR relationship, the framework identifies the trigonal MST phase as pivotal to phase-change behavior, opening new opportunities for exploration in this complex materials system.

## 5. Materials and Methods

**Sample preparation**
We deposited ternary thin film composition spreads on 3 inch Si/$SiO_2$ substrates using an ultrahigh vacuum (base pressure: 2.7 × $10^{-6}$ Pa (2 × $10^{-8}$ Torr)) magnetron sputtering system (AJA Orion-3) at room temperature using a Si mask placed in contact with the substrate to delineate 177 individual compositions evenly with a 4.5 mm separation. High purity Mn (99.95 %), Sb (99.99 %) and Te (99.99 %) targets (Kurt J. Lesker Co.) were used to co-sputter in ultrahigh purity Argon (99.9997 %, Airgas) at a pressure of 0.61 Pa (4.6 × $10^{-3}$ Torr). The thin film spread was deposited over 30 min using 40 W radio frequency (RF), 13 W DC and 25 W RF power sources for Mn, Sb and Te, respectively. The co-sputtering results in a minimal thickness gradient (100 nm $\pm$ 10 nm), measured by profilometer. The thin films were capped with 30 nm $SiO_2$ to avoid surface contamination. We obtained the crystalline spread annealing as-deposited spread on a hotplate in an $N_2$ glove box for 20 min. Chemical composition of Mn-Sb-Te thin film library is determined by using wavelength dispersive spectroscopy in an electron probe microanalyzer (JXA 8900R Microprobe) prior to structural and functional property characterization, with an acceleration voltage of 15 kV. Calibration was done using polished pure metal with an experimental error margin of atomic concentration of < 0.3 %. We assumed chemical stoichiometry is homogeneous within 4 mm, which is defined as one compositional data point.

**Structural Chracterization**
XRD was carried out on a crystalline thin-film library. A Discover powder diffractometer (Bruker C2/D8) of Cu-$K_\alpha$ radiation with a high-temperature stage was used to collect XRD images. The exposure time was 5 min for each frame, and 3 frames were done for one diffraction pattern. That is, each measurement takes 15 min in total. The diffraction was integrated into 1D data with the $2\theta$ range from 12° to 56° with a 0.05° step size automatically using home-built codes.

**Electrical resistance measurement**

Electrical contact measurements were performed on an amorphous thin-film composition library using an automated probe station (Signatone CM-250) equipped with two-point probes. The electrical response was recorded via a digital multimeter (Keithley 2000) under a constant current of $1 \times 10^{-5}$ A, and an integration time of 1 s. To ensure reproducibility, three independent contact measurements were conducted at each composition point, each lasting approximately 30 s, and the average resistance value was used for subsequent analysis. Thus, each measurement takes 1.5 min.

**Algorithm**

XRD patterns were first processed for background removal using the *pybaselines* package. To decompose the XRD dataset into representative structural components, NMF was performed using the *scikit-learn* library, employing random initialization and a fixed random seed for reproducibility, with convergence achieved within 3000 iterations. Note that for the ternary system, 7 components were selected to represent the possible phase configurations (3 elemental, 3 binary, and 1 ternary) reflecting the anticipated complexity of the multi-phase structure. Both functional and structural properties were independently standardized using the *StandardScaler* from *scikit-learn*, such that every property was normalized only by its own mean and standard deviation. Model training and uncertainty estimation were performed in the standardized space. Predictions were capped to the physical range $[0,1]$ when applicable, and subsequently inverse-transformed back to the original units, with predictive standard deviations rescaled accordingly. This imposed equal weighting across each input dimension. Multi-output GP regression was performed using the *GPy* library with a linear coregionalization model (LCM). Inputs were two-dimensional Cartesian coordinates derived from ternary compositions, and outputs comprised 7 structural components from NMF and 1 functional property. A single latent function (weight rank = 1) with a Matérn 3/2 kernel captured correlations across outputs. All hyperparameters were optimized over 5000 iterations. For comparison with the multi-output LCM, an independent GP regression was implemented using the intrinsic coregionalization model (ICM). In this configuration, the off-diagonal weights of coregionalization matrix B were fixed to zero, effectively decoupling the outputs and treating each as independent, while the scaling parameters for each output were left unconstrained. For each output, the GP provided the predictive mean and uncertainty. The predicted 7 phase components were combined with the previously extracted NMF abundances to reconstruct the XRD patterns, allowing the reconstructed patterns to be quantitatively compared with the experimental measurements. Piecewise clustering comparison was performed with $M = 5$ groups using spectral clustering, with each diffraction pattern was encoded as a feature vector, and an affinity matrix was constructed using cosine similarity. The number of clusters, $M$, was determined using the elbow method. This approach enabled identification of phase boundaries and facilitated comparison with NMF-derived phases. Phase identification was carried out by comparison with reference patterns from the database Inorganic Crystal Structure Database (ICSD), FIZ Karlsruhe, Karlsruhe, Germany.

## Acknowledgement

I.T. and C.Ri.O acknowledge funding provided by the National Science Foundation (awards DMR-2329087/2329088), supported in part by industry partners, as specified in the Future of Semiconductors (FuSe) program. The algorithm development was supported by 3DFeM2, an EFRC funded by the U.S. DOE, Office of Science, Basic Energy Sciences under Award Number DE-SC0021118. This work is also partly funded by DTRA through grant number HDTRA12410015. Some of the materials synthesis was performed at NanoCenter, University of Maryland.

## References

1. Liang, H. *et al.* Real-time experiment-theory closed-loop interaction for autonomous materials science. *Science Advances* **11**, eadu7426 (2025).
2. Dai, T. *et al.* Autonomous mobile robots for exploratory synthetic chemistry. *Nature* **635**, 890–897 (2024).
3. Dave, A. *et al.* Autonomous optimization of non-aqueous Li-ion battery electrolytes via robotic experimentation and machine learning coupling. *Nat Commun* **13**, 5454 (2022).
4. Wang, C. *et al.* Autonomous platform for solution processing of electronic polymers. *Nat Commun* **16**, 1498 (2025).
5. McDannald, A. *et al.* On-the-fly autonomous control of neutron diffraction via physics-informed Bayesian active learning. *Appl. Phys. Rev.* **9**, 021408 (2022).
6. Martin, T. B., Sutherland, D. R., McDannald, A., Kusne, A. G. & Beaucage, P. A. Autonomous Small-Angle Scattering for Accelerated Soft Material Formulation Optimization. *Chem. Mater.* **37**, 4272–4281 (2025).

7. Ament, S. *et al.* Autonomous materials synthesis via hierarchical active learning of nonequilibrium phase diagrams. *Science Advances* **7**, eabg4930 (2021).

8. Gregoire, J. M., Zhou, L. & Haber, J. A. Combinatorial synthesis for AI-driven materials discovery. *Nat. Synth* **2**, 493–504 (2023).

9. Takeuchi, I., Lauterbach, J. & Fasolka, M. J. Combinatorial materials synthesis. *Materials Today* **8**, 18–26 (2005).

10. Stach, E. *et al.* Autonomous experimentation systems for materials development: A community perspective. *Matter* **4**, 2702–2726 (2021).

11. Montoya, J. H. *et al.* Autonomous intelligent agents for accelerated materials discovery. *Chem. Sci.* **11**, 8517–8532 (2020).

12. Gomes, C. P. *et al.* CRYSTAL: a multi-agent AI system for automated mapping of materials' crystal structures. *MRS Communications* **9**, 600–608 (2019).

13. Kusne, A. G. & McDannald, A. Scalable multi-agent lab framework for lab optimization. *Matter* **6**, 1880–1893 (2023).

14. Kusne, A. & Mcdannald, A. Agent, Agentic, and Distributed Artificial Intelligence: From Managing Next-Generation Labs to the Philosophy of Science. https://www.authorea.com/users/280747/articles/1340689-agent-agentic-and-distributed-artificial-intelligence-from-managing-next-generation-labs-to-the-philosophy-of-science.

15. Roch, L. M. *et al.* ChemOS: Orchestrating autonomous experimentation. *Science Robotics* **3**, eaat5559 (2018).

16. Rahmanian, F. *et al.* Enabling Modular Autonomous Feedback-Loops in Materials Science through Hierarchical Experimental Laboratory Automation and Orchestration. *Advanced Materials Interfaces* **9**, 2101987 (2022).

17. Guevarra, D. *et al.* Orchestrating nimble experiments across interconnected labs. *Digital Discovery* **2**, 1806–1812 (2023).

18. Pendleton, I. M. *et al.* Experiment Specification, Capture and Laboratory Automation Technology (ESCALATE): a software pipeline for automated chemical experimentation and data management. *MRS Communications* **9**, 846–859 (2019).

19. Sim, M. *et al.* ChemOS 2.0: An orchestration architecture for chemical self-driving laboratories. *Matter* **7**, 2959–2977 (2024).

20. Beaucage, P., Sutherland, D. & Martin, T. Dueling Robots: Concurrent, Robotic, AI-Driven SAXS and SANS to Solve Industrial Problems. *Struct. Dyn.* **12**, A190 (2025).

21. Bonilla, E., Chai, K. M. A. & Williams, C. K. I. Multi-Task Gaussian process prediction. *Proc. Adv. Neural Inf. Process. Syst* **20**, 153–160 (2008).

22. Alvi, S. M. A. A. *et al.* Hierarchical Gaussian process-based Bayesian optimization for materials discovery in high entropy alloy spaces. *Acta Materialia* **289**, 120908 (2025).

23. Khatamsaz, D., Vela, B. & Arróyave, R. Multi-objective Bayesian alloy design using multi-task Gaussian processes. *Materials Letters* **351**, 135067 (2023).

24. Gilad Kusne, A., McDannald, A. & DeCost, B. Learning material synthesis-process-structure-property relationship by data fusion: Bayesian Coregionalization N-Dimensional Piecewise Function Learning. *arXiv e-prints* arXiv:2311.06228 (2023) doi:10.48550/arXiv.2311.06228.

25. N. Slautin, B. *et al.* Co-orchestration of multiple instruments to uncover structure–property relationships in combinatorial libraries. *Digital Discovery* **3**, 1602–1611 (2024).

26. Kusne, A. G. *et al.* On-the-fly closed-loop materials discovery via Bayesian active learning. *Nat Commun* **11**, 5966 (2020).

27. Xiang, X.-D. *et al.* A Combinatorial Approach to Materials Discovery. *Science* **268**, 1738–1740 (1995).

28. Szymanski, N. J. *et al.* Adaptively driven X-ray diffraction guided by machine learning for autonomous phase identification. *npj Comput Mater* **9**, 31 (2023).

29. Maffettone, P. M. *et al.* Crystallography companion agent for high-throughput materials discovery. *Nat Comput Sci* **1**, 290–297 (2021).

30. Szymanski, N. J., Bartel, C. J., Zeng, Y., Tu, Q. & Ceder, G. Probabilistic Deep Learning Approach to Automate the Interpretation of Multi-phase Diffraction Spectra. *Chem. Mater.* **33**, 4204–4215 (2021).

31. Long, C. J. *et al.* Rapid structural mapping of ternary metallic alloy systems using the combinatorial approach and cluster analysis. *Rev. Sci. Instrum.* **78**, 072217 (2007).

32. Stanev, V. *et al.* Unsupervised phase mapping of X-ray diffraction data by nonnegative matrix factorization integrated with custom clustering. *npj Comput Mater* **4**, 43 (2018).

33. Long, C. J., Bunker, D., Li, X., Karen, V. L. & Takeuchi, I. Rapid identification of structural phases in combinatorial thin-film libraries using x-ray diffraction and non-negative matrix factorization. *Rev Sci Instrum* **80**, 103902 (2009).

34. Suram, S. K. *et al.* Automated Phase Mapping with AgileFD and its Application to Light Absorber Discovery in the V–Mn–Nb Oxide System. *ACS Comb. Sci.* **19**, 37–46 (2017).

35. Chen, D. *et al.* Automating crystal-structure phase mapping by combining deep learning with constraint reasoning. *Nat Mach Intell* **3**, 812–822 (2021).

36. Maffettone, P. M., Daly, A. C. & Olds, D. Constrained non-negative matrix factorization enabling real-time insights of in situ and high-throughput experiments. *Appl. Phys. Rev.* **8**, 041410 (2021).

37. Mudgal, M. *et al.* Magnetotransport and electronic structure of the axion insulator ${\mathrm{MnSb}}_{8}{\mathrm{Te}}_{13}$. *Phys. Rev. B* **110**, 045124 (2024).

38. Xi, M. *et al.* Relationship between Antisite Defects, Magnetism, and Band Topology in MnSb2Te4 Crystals with TC ≈ 40 K. *J Phys Chem Lett* **13**, 10897–10904 (2022).

39. Mn-Rich MnSb2Te4: A Topological Insulator with Magnetic Gap Closing at High Curie Temperatures of 45–50 K - Wimmer - 2021 - Advanced Materials - Wiley Online Library. https://advanced.onlinelibrary.wiley.com/doi/full/10.1002/adma.202102935.

40. Riberolles, S. X. M. *et al.* Evolution of magnetic interactions in Sb-substituted ${\mathrm{MnBi}}_{2}{\mathrm{Te}}_{4}$. *Phys. Rev. B* **104**, 064401 (2021).

41. Adam, A. A. E., Cheng, X., Guan, X. & Miao, X. Ferromagnetism modulation by phase change in Mn-doped GeTe chalcogenide magnetic materials. *Appl. Phys. A* **117**, 2115–2119 (2014).

42. Song, W.-D., Shi, L.-P. & Chong, T.-C. Magnetic Properties and Phase Change Features in Fe-Doped Ge–Sb–Te. *Journal of Nanoscience and Nanotechnology* **11**, 2648–2651 (2011).

43. Ding, D. *et al.* Origin of ferromagnetism and the design principle in phase-change magnetic materials. *Phys. Rev. B* **84**, 214416 (2011).

44. Liu, J. Toward flexible memory application: high-performance phase-change magnetic material Fe:GeTe films realized via quasi-van der Waals epitaxy. *J. Mater. Chem. C* **10**, 9891–9901 (2022).

45. Raoux, S., Xiong, F., Wuttig, M. & Pop, E. Phase change materials and phase change memory. *MRS Bulletin* **39**, 703–710 (2014).

46. Gu, R. *et al.* Stretched non-negative matrix factorization. *npj Comput Mater* **10**, 193 (2024).

47. Kusne, A. G., McDannald, A. & DeCost, B. Learning material synthesis–process–structure–property relationship by data fusion: Bayesian co-regionalization N-dimensional piecewise function learning. *Digital Discovery* **3**, 2211–2225 (2024).

48. Iwasaki, Y., Kusne, A. G. & Takeuchi, I. Comparison of dissimilarity measures for cluster analysis of X-ray diffraction data from combinatorial libraries. *npj Comput Mater* **3**, 4 (2017).

49. Utimula, K. *et al.* Machine-Learning Clustering Technique Applied to Powder X-Ray Diffraction Patterns to Distinguish Compositions of ThMn12-Type Alloys. *Advanced Theory and Simulations* **3**, 2000039 (2020).

50. Forrester, C. R. *et al*. Structural and magnetic properties of molecular beam epitaxy (MnSb2Te4)x(Sb2Te3)1−x topological materials with exceedingly high Curie temperature. *APL Materials* **12**, 071109 (2024).

51. Liu, Y. *et al*. Site Mixing for Engineering Magnetic Topological Insulators. *Phys. Rev. X* **11**, 021033 (2021).

52. Saxena, A. & Awana, V. P. S. Growth and characterization of the magnetic topological insulator candidate Mn2Sb2Te5. *J. Phys.: Condens. Matter* **36**, 085704 (2023).

53. Imai, H. Shift Invariance Property of a Non-Negative Matrix Factorization. *IEICE Trans. Fundamentals* **E103.A**, 580–581 (2020).

54. Hoyer, P. O. & Hoyer, P. Non-negative Matrix Factorization with Sparseness Constraints.

55. Wang, Z. & Min, W. Graph Regularized NMF with L20-norm for Unsupervised Feature Learning. Preprint at https://doi.org/10.48550/arXiv.2403.10910 (2024).

56. Brouwer, T., Frellsen, J. & Lió, P. Comparative Study of Inference Methods for Bayesian Nonnegative Matrix Factorisation. Preprint at https://doi.org/10.48550/arXiv.1707.05147 (2017).

57. Lu, J. & Chai, C. P. Robust Bayesian Nonnegative Matrix Factorization with Implicit Regularizers. Preprint at https://doi.org/10.48550/arXiv.2208.10053 (2022).

58. Ozbayram, O., Olivier, A. & Graham-Brady, L. Heteroscedastic Gaussian Process Regression for material structure–property relationship modeling. *Computer Methods in Applied Mechanics and Engineering* **431**, 117326 (2024).

59. Le, Q. V., Smola, A. J. & Canu, S. Heteroscedastic Gaussian process regression. in *Proceedings of the 22nd international conference on Machine learning - ICML '05* 489–496 (ACM Press, Bonn, Germany, 2005). doi:10.1145/1102351.1102413.

60. Zhou, F. *et al.* Heterogeneous Multi-Task Gaussian Cox Processes. Preprint at https://doi.org/10.48550/arXiv.2308.15364 (2023).

61. Moreno-Muñoz, P., Artés, A. & Álvarez, M. Heterogeneous Multi-output Gaussian Process Prediction.

62. Jiang, X., Georgaka, S., Rattray, M. & Álvarez, M. A. Scalable Multi-Output Gaussian Processes with Stochastic Variational Inference. Preprint at https://doi.org/10.48550/arXiv.2407.02476 (2025).

63. Lupoiu, R. *et al.* A multi-agentic framework for real-time, autonomous freeform metasurface design. *Science Advances* **11**, eadx8006 (2025).

64. Ghafarollahi, A. & Buehler, M. J. Autonomous Inorganic Materials Discovery via Multi-Agent Physics-Aware Scientific Reasoning. Preprint at https://doi.org/10.48550/arXiv.2508.02956 (2025).